\documentclass[10pt,preprint]{emulateapj}
\usepackage{graphicx}

\shorttitle{The JCMT Transient Survey}
\shortauthors{Herczeg et al.}

\begin{document}

\title{How do stars gain their mass?  A JCMT/SCUBA-2 Transient Survey
  of Protostars in Nearby Star Forming Regions}

\author{
Gregory J. Herczeg\altaffilmark{1},
Doug Johnstone\altaffilmark{2,3},
Steve Mairs\altaffilmark{2,3},
Jennifer Hatchell\altaffilmark{4},
Jeong-Eun Lee\altaffilmark{5},
Geoffrey C. Bower\altaffilmark{6},
Huei-Ru Vivien Chen\altaffilmark{7},
Yuri Aikawa\altaffilmark{8},
Hyunju Yoo\altaffilmark{9,5},
Sung-Ju Kang\altaffilmark{10},
Miju Kang\altaffilmark{10},
Wen-Ping Chen\altaffilmark{11},
Jonathan P. Williams\altaffilmark{12},
Jaehan Bae\altaffilmark{13},
Michael M. Dunham\altaffilmark{14,15},
Eduard I. Vorobiov\altaffilmark{16,17,18},
Zhaohuan Zhu\altaffilmark{19},
Ramprasad Rao\altaffilmark{20},
Helen Kirk\altaffilmark{2},
Satoko Takahashi\altaffilmark{21,22},
Oscar Morata\altaffilmark{20},
Kevin Lacaille\altaffilmark{23},
James Lane\altaffilmark{3},
Andy Pon\altaffilmark{24},
Aleks Scholz\altaffilmark{25},
Manash R. Samal\altaffilmark{11},
Graham S. Bell\altaffilmark{26},
Sarah Graves\altaffilmark{26},
E'lisa M. Lee\altaffilmark{26},
Harriet Parsons\altaffilmark{26},
Yuxin He\altaffilmark{27},
Jianjun Zhou\altaffilmark{27},
Mi-Ryang Kim\altaffilmark{28},
Scott Chapman\altaffilmark{23},
Emily Drabek-Maunder\altaffilmark{29},
Eun Jung Chung\altaffilmark{10},
Stewart P. S. Eyres\altaffilmark{30},
Jan Forbrich \altaffilmark{31,15},
Lynne A. Hillenbrand\altaffilmark{32},
Shu-ichiro Inutsuka\altaffilmark{33},
Gwanjeong Kim\altaffilmark{10},
Kyoung Hee Kim\altaffilmark{34},
Yi-Jehng Kuan\altaffilmark{35,20},
Woojin Kwon\altaffilmark{10,36},
Shih-Ping Lai\altaffilmark{7,20},
Bhavana Lalchand\altaffilmark{11},
Chang Won Lee\altaffilmark{10,36},
Chin-Fei Lee\altaffilmark{20}
Feng Long\altaffilmark{1,37},
A-Ran Lyo\altaffilmark{10},
Lei Qian\altaffilmark{38},
Peter Scicluna\altaffilmark{20},
Archana Soam\altaffilmark{10},
Dimitris Stamatellos\altaffilmark{30},
Shigehisa Takakuwa\altaffilmark{39},
Ya-Wen Tang\altaffilmark{20},
Hongchi Wang\altaffilmark{40},
Yiren Wang\altaffilmark{1,37}}

\altaffiltext{1}{Kavli Institute for Astronomy and Astrophysics, Peking University, Yiheyuan 5, Haidian Qu, 100871 Beijing, People's Republic of China}

\altaffiltext{2}{NRC Herzberg Astronomy and Astrophysics, 5071 West Saanich Rd, Victoria, BC, V9E 2E7, Canada}

\altaffiltext{3}{Department of Physics and Astronomy, University of Victoria, Victoria, BC, V8P 5C2, Canada}

\altaffiltext{4}{Physics and Astronomy, University of Exeter, Stocker Road, Exeter EX4 4QL, UK}

\altaffiltext{5}{School of Space Research, Kyung Hee University, 1732, Deogyeong-Daero, Giheung-gu Yongin-shi, Gyunggi-do 17104, Korea}

\altaffiltext{6}{Academia Sinica Institute of Astronomy and Astrophysics, 645 N. A'ohoku Place, Hilo, HI 96720, USA}

\altaffiltext{7}{Department of Physics and Institute of Astronomy, National Tsing Hua University, Taiwan}

\altaffiltext{8}{Department of Astronomy, University of Tokyo, Tokyo, Japan}

\altaffiltext{9}{Chungnam National University, Korea}

\altaffiltext{10}{Korea Astronomy and Space Science Institute, 776 Daedeokdae-ro, Yuseong-gu, Daejeon 34055, Republic of Korea}

\altaffiltext{11}{Graduate Institute of Astronomy, National Central University, 300 Jhongda Road, Zhongli, Taoyuan, Taiwan}

\altaffiltext{12}{Institute for Astronomy, University of Hawaii at Manoa, Honolulu, HI 96822, USA}

\altaffiltext{13}{Department of Astronomy, University of Michigan, 1085 S. University Ave., Ann Arbor, MI 48109, USA}

\altaffiltext{14}{Department of Physics, The State University of New York at Fredonia, Fredonia, NY 14063, USA}

\altaffiltext{15}{Harvard-Smithsonian Center for Astrophysics, 60 Garden Street, Cambridge, MA 02138, USA}

\altaffiltext{16}{Institute of Fluid Mechanics and Heat Transfer, TU Wien, Vienna, 1060, Austria}

\altaffiltext{17}{Research Institute of Physics, Southern Federal University, Stachki Ave. 194, Rostov-on-Don, 344090, Russia}

\altaffiltext{18}{University of Vienna, Department of Astrophysics, Vienna, 1180, Austria}

\altaffiltext{19}{Physics and Astronomy Department, University of Nevada, Las Vegas}

\altaffiltext{20}{Academia Sinica Institute of Astronomy and Astrophysics, P. O. Box 23-141, Taipei 10617, Taiwan}

\altaffiltext{21}{Joint ALMA Observatory, Alonso de C{\'{o}}rdova 3107, Vitacura, Santiago, Chile}
\altaffiltext{22}{National Astronomical Observatory of Japa, 2-21-1 Osawa, Mitaka, Tokyo 181-8588, Japan}

\altaffiltext{23}{Department of Physics and Atmospheric Science, Dalhousie University, Halifax, NS, B3H 4R2, Canada}

\altaffiltext{24}{Department of Physics and Astronomy, The University of Western Ontario, 1151 Richmond Street, London, N6A 3K7, Canada}

\altaffiltext{25}{SUPA, School of Physics \& Astronomy, North Haugh, St Andrews, KY16 9SS, United Kingdom}

\altaffiltext{26}{East Asian Observatory, 660 N.\ A`oh\=ok\=u Place, Hilo, HI 96720, USA}

\altaffiltext{27}{Xinjiang Astronomical Observatory, Chinese Academy of Sciences, Urumqi, China}

\altaffiltext{28}{Department of Physics, Institute for Astrophysics, Chungbuk National University, Republic of Korea}

\altaffiltext{29}{Cardiff University, School of Physics and Astronomy, The Parade, Cardiff, CF24 3AA}

\altaffiltext{30}{Jeremiah Horrocks Institute for Mathematics, Physics \& Astronomy, University of Central Lancashire, Preston, PR1 2HE, UK}

\altaffiltext{31}{Centre for Astrophysics Research, School of Physics, Astronomy and Mathematics, University of Hertfordshire, College Lane, Hatfield AL10 9AB, UK}

\altaffiltext{32}{Department of Astronomy; MC 249-17, California Institute of Technology, Pasadena, CA 91125, USA}

\altaffiltext{33}{Department of Physics, Graduate School of Science, Nagoya University, 464-8602, Nagoya, Japan}

\altaffiltext{34}{Korea National University of Education,  Taeseongtabyeon-ro, Grangnae-myeon, Heungdeok-gu, Cheongju-si, Chungbuk 28173, Korea}

\altaffiltext{35}{Department of Earth Sciences, National Taiwan Normal University, Taipei 116, Taiwan}

\altaffiltext{36}{Korea University of Science and Technology, 217 Gajang-ro, Yuseong-gu, Daejeon 34113, Republic of Korea}

\altaffiltext{37}{Department of Astronomy, Peking University, Yiheyuan 5, Haidian Qu, 100871 Beijing, People's Republic of China}

\altaffiltext{38}{National Astronomical Observatories, Chinese Academy of Sciences, Beijing, China}

\altaffiltext{39}{Department of Physics and Astronomy, Graduate School of Science and Engineering, Kagoshima University, 1-21-35 Korimoto, Kagoshima, Kagoshima 890-0065, Japan}

\altaffiltext{40}{Purple Mountain Observatory, \& Key Laboratory for Radio Astronomy, Chinese Academy of Sciences, 2 West Beijing Road, Nanjing 210008, China}

\begin{abstract}
Most protostars have luminosities that are fainter than expected
from steady accretion over the protostellar lifetime.   The solution
to this problem may  lie in episodic mass accretion -- prolonged periods of very low accretion punctuated
by short bursts of rapid accretion. 
However, the timescale and
amplitude for variability at the protostellar phase is almost entirely
unconstrained.  In {\it A JCMT/SCUBA-2 Transient Survey of Protostars in Nearby Star Forming Regions}, we are monitoring monthly with SCUBA-2 the sub-mm emission in eight fields within nearby ($<500$ pc) star forming regions to measure
the accretion variability of protostars.  The total survey area of $\sim 1.6$ sq.deg. includes $\sim 105$ peaks with peaks brighter than 0.5 Jy/beam (43 associated with embedded protostars or disks) and 237 peaks of 0.125--0.5 Jy/beam (50 with embedded protostars or disks).  Each field has enough bright peaks
for flux calibration relative to other peaks in the same field, which improves upon the nominal flux calibration uncertainties of sub-mm observations to reach a precision of $\sim 2-3$\% rms, and also provides quantified confidence in 
any measured variability.  
The timescales and amplitudes of
any sub-mm variation will then be converted into variations in accretion rate and subsequently used to infer the physical causes of the variability.  
This survey is the first dedicated survey for
sub-mm variability and complements other transient surveys at optical
and near-IR wavelengths, which are not sensitive to accretion variability of deeply embedded protostars.
\end{abstract}
 




\section{INTRODUCTION}
Low-mass stars form through the gravitational collapse of molecular cloud cores. The evolution
of mass accretion onto a forming protostar depends on the rate at which the interior of the core
collapses, the role of the circumstellar disk as a temporary mass reservoir and transportation
mechanism, and the physics of how the inner disk accretes onto the
stellar surface.
In dynamical models
of gravitational collapse of a spherical protostellar core \citep[e.g.][]{shu77,shu87,masunaga00}, the young
star grows steadily from the infalling envelope at a
rate of a few $10^{-6}$ M$_\odot$ yr$^{-1}$. However, \citet{kenyon90} found that
luminosities of most protostars fall far below those expected from
energy release by steady accretion over protostellar lifetimes.  This
luminosity problem has since been confirmed with an improved census
of protostars (e.g., \citealt{enoch09,dunham10}; see also discussions in \citealt{dunham14} and \citealt{fischer17}).

The resolution of the luminosity problem likely requires a either a
time-dependent or mass-dependent accretion rate (e.g., discussions in \citealt{hartmann16} and \citealt{fischer17}; see also, e.g., ~\citealt{offner11,myers12})
  Observationally, strong but mostly indirect
evidence suggests that the accretion rate is punctuated by short
bursts of rapid accretion, often termed {\it episodic accretion}
\citep{kenyon90,dunham10,dunham12}.
The form of this time dependence may have a
lasting affect on the evolution of stars \citep[e.g.][]{hartmann97,baraffe10,baraffe17,vorobyov17age}, the
physical structure of disks and their propensity to fragment, 
\citep[e.g.][]{stamatellos11,stamatellos12,vorobyov14arep}, and the chemistry of disks
and envelopes \citep[e.g.][]{kim12,jorgensen13,vorobyov13,visser15,harsono15,owen15,cieza16}.

The suggestion of accretion bursts in protostars has 
significant support from later stages of pre-main sequence stellar
evolution. Spectacular outbursts\footnote{The classification scheme of EXor and FUor outbursts is sometimes vague, with both classes likely including diverse phenomenon.} with optical brightness increases of
$\sim 5$ mag are interpreted as 
accretion rate increases of 2-4 orders of
magnitude and can last for months \citep[called EXors following the
prototype EX Lup, e.g.][]{herbig08,aspin10}
or decades \citep[called FUors following the prototype FU Ori, e.g.][]{herbig77,hartmann96}.  Because most transient searches use optical
photometry, these accretion outbursts are detected only on young stellar objects that are
optically bright and are therefore biased to variability at or near the end of their main phase of stellar growth.
Only a few outbursts have been detected on a deeply embedded Class 0 star
\citep{kospal07,safron15,hunter17},
the stage when the star should accrete much of its mass -- although many FUor objects retain some envelopes and are classified as Class I objects
\citep[e.g.][]{zhu08,caratti11,caratti16,fischer12,green13,kospal17}.  
Indirect evidence for outbursts includes chemical
signatures of past epochs with high luminosity \citep[e.g.][]{kim12,vorobyov13,jorgensen15,frimann17}
and periodic shocks/bullets along protostellar jets,
which may offer a historical record of accretion events
\citep[e.g.][]{reipurth89,raga02,plunkett15}.  In addition to these large events, instabilities in the inner disk likely lead to more frequent but smaller bursts of accretion, as seen in more evolved disks \citep[e.g.][]{costigan14,venuti14,cody17}.

Directly observing either large outbursts or accretion flickers on protostars is challenging
because they are deeply embedded in dense envelopes.  The
accretion luminosity is not directly visible to us, and is instead absorbed by the envelope and reprocessed into photons with lower energies, which then escape from the system.  
Models of an accretion burst indicate that the enhanced accretion
luminosity heats dust in the envelope  \citep{johnstone13}.  The dust
is then seen as brighter emission at far-IR through sub-mm
wavelengths.  The change in luminosity is strongest at far-IR
wavelengths, which traces the effecitive photosphere of the envelope,
where the envelope becomes transparent when the local temperature
drops below $\sim 100$ K.  Single-dish observations at sub-mm wavelengths have
large scales, which tends to probe the temperature structure of the outer
envelope.  When the protostellar luminosity increases, the outer
envelope is expected to become hotter.  Since the atmosphere
of the Earth is opaque in the far-IR, and the most heavily embedded
objects are not visible at optical/near-IR wavelengths, 
sub-mm observations provide us with our best ground-based window into the
protostar -- a snapshot of the accretion rate, averaged over the
timescale of a few weeks for the luminosity burst to propagate through the envelope.
While some far-IR/sub-mm variability has
indeed been detected on protostars, these detections are mostly based
on transients identified in optical/near-IR surveys and have few
epochs of flux measurements at far-IR/sub-mm wavelengths
\citep[e.g.][]{billot12,scholz13var,balog14,safron15,onozato15}.

 In this paper, we describe our novel James Clerk Maxwell Telescope (JCMT) survey, {\it A JCMT/SCUBA-2 Transient Survey
  of Protostars in Nearby Star Forming Regions}, shortened to {\it JCMT-Transient}, in
 which we use Submillimetre Common-User Bolometer Array 2 
 \citep[SCUBA-2;][]{holland13}
to monitor the sub-mm flux of deeply embedded
 protostars in eight fields within nearby star forming regions.
  This
 is the first dedicated long-term sub-mm monitoring program.  The only previous sub-mm monitoring programs probed variability in synchrotron radiation from the Sagittarius A* at the galactic center over five consecutive nights \citep{haubois12} and the black hole X-ray binary V404 Cyg over $4.5$ hrs \citep{tetarenko17}.
Although large outbursts, with a factor of 100 increases in source luminosity, are rare \citep{scholz13var,hillenbrand15}, our survey should also reveal the lower-amplitude variability (with changes of a factor or $<10$ in luminosity) that are commonly detected on classical T Tauri stars.
 In \S 2 we
 describe our observational plans.  In \S 3 we describe initial
 results, including the stability of our flux calibration.  In \S 4,
 we discuss the expected contributions of this survey to our understanding of protostellar
 variability and related applications for this dataset.  In \S 5 we discuss ancillary science related to disks, VeLLOs, filaments, and non-thermal emission.  In \S 6 we discuss our expectations for the future results from this program.

\section{OVERVIEW OF SURVEY METHODOLOGY}

Our ongoing JCMT survey program, M16AL001, consists of monitoring 450 and 850 $\mu$m emission from eight young regions
that  are rich in protostars, as identified in previous Spitzer,
Herschel, and SCUBA-2 Gould Belt Surveys.
Sub-mm monitoring surveys have been challenging in the past because of
calibration uncertainties.  The wide SCUBA-2 field-of-view allows us to use multiple bright sources in the same field to calibrate the image relative to other bright objects in the field, the sub-mm equivalent of differential photometry.

\subsection{Sample Selection}

We selected eight 30$^\prime$ regions of nearby ($<500$ pc), active
star formation to maximize the number of protostars and
disks in the fields, with a preference to the youngest regions while also avoiding regions with the most complex, confused features.  The fields include a total of 182 Class 0/I objects, 132 flat-spectrum objects, and 670 disks (see Table~\ref{tab:sources}), as
previously classified from Spitzer SEDs by \citet{dunham10}, \citet{stutz13}, and \citet{megeath16}.  Each
region includes 3--41 peaks with 850 $\mu$m fluxes above 0.5 Jy/beam and
12--120 peaks above 0.12 Jy/beam, and 3--14 protostars associated with those peaks.  All
requested fields have a past epoch from the JCMT Gould Belt
Survey \citep{ward-thompson07}, along with complementary Spitzer
mid-IR \citep{dunham15,megeath16} and Herschel far-IR
imaging \citep{andre14}.

\begin{table*}[!t]
\begin{center}
\caption{Description of the eight selected fields}
\label{tab:sources}
\begin{tabular}{llc|cccc|ccc|cc|cc}
\hline
Name    &  Location       &     dist  & 
\multicolumn{4}{c}{\# of Peaks$^a$} 
    &          \multicolumn{3}{c}{Spitzer Sources$^b$} & 
    \multicolumn{2}{c}{Class 0/I$^c$} & 
    \multicolumn{2}{c}{Disks$^c$} \\
    & & pc  & $>0.125$ & $>0.25$ &  $>0.5$ & $>1.0$ & 0/I & Flat & II & $>0.125$ & $>0.5$ & $>0.125$ & $>0.5$ \\
    \hline
Perseus - NGC 1333 & 032854+311652 & 270$^d$ & 33 & 24 & 10 & 6 & 34 & 14 & 62 & 14 & 6 & 1 & 0\\
Perseus - IC348 & 034418+320459 & 303$^d$       & 12 & 4 & 3 & 2 & 13 & 8 & 114 & 5 & 3 & 0 & 0\\
Orion A - OMC2/3 & 053531-050038 & 388$^e$    & 120 & 77 & 41 & 25 & 64 & -- & 600 & 12 & 10 & 12 & 4\\
Orion B - NGC2024 & 054141-015351 & 423$^e$    &38 & 14 & 8 & 4 & 26 & -- & 232 & 3 & 2 & 1 & 0\\
Orion B - NGC2068 & 054613-000605 & 388$^e$  &31 & 24 & 12 & 5 & 22 & -- & 117 & 9 & 6 & 0 & 0\\
Ophiuchus & 162705-243237     & 137$^f$              & 41 & 23 & 7 & 3 & 21 & 30 & 87 & 12 & 1 & 6 & 2\\
Serpens Main & 182949+011520 & 436$^g$              & 16 & 15 & 10 & 7 & 19 & 9 & 51 & 8 & 7 & 0 & 0\\
Serpens South & 183002-020248 & 436$^g$             & 51 & 27 & 14 & 2 & 50 & 34 & 142 & 10 & 1 & 1 & 0\\
\hline
\multicolumn{14}{l}{$^a$\# of individual peaks with SCUBA-2 850 $\mu$m flux brighter than listed.}\\
\multicolumn{14}{l}{$^b$Total \# of protostars in field, as identified by \citet{stutz13}, \citet{dunham15}, and \citet{megeath16}.}\\
\multicolumn{14}{l}{$^c$Protostars and disks located within $7^{\prime\prime}$ of a peak, see also Appendix A}\\
\multicolumn{14}{l}{$^d$This work using parallaxes from the Gaia DR1 TGAS catalogue.}\\
\multicolumn{14}{l}{$^e$Parallaxes from the VLBI GOBELINS program \citep{kounkel17}.}\\
\multicolumn{14}{l}{$^f$Parallax from the VLBI GOBELINS program \citep{ortiz17oph}.}\\
\multicolumn{14}{l}{$^g$Parallaxes from the VLBI GOBELINS program \citep{ortiz17ser}.}\\
\end{tabular}
\end{center}
\end{table*}

\subsection{Observing Strategy}

The SCUBA-2 instrument is a 10,000 pixel bolometer camera that
images simultaneously at 450 and 850 $\mu$m with $9\farcs8$ and $14\farcs6$ 
resolution \citep{holland13,dempsey13}.  Both focal planes consist of four
subarrays of 1280 bolometers that simultaneously cover a field with an
$\sim 8^{\prime}$ diameter.  The regions are observed in a 
pong $1800^{\prime\prime}$ pattern, in which SCUBA-2 scans over a field-of-view of $30^\prime$
diameter to produce an image with smooth sensitivity across the map \citep{kackley10}.

Our observations are being obtained in weather bands 1--3, which
correspond to different levels of atmospheric H$_2$O column densities
that lead to opacities of $\tau<0.12$ at 225 GHz.  
\citet{mairs17} provides a complete list of observations obtained
through February 2017, including $\tau$ and sensitivity.  To date,
21\% of our observations have been obtained in Band 1 (the driest
weather, $\tau<0.05$ at 225 GHz) and 39\% have been observed in Band 2
($0.05<\tau<0.08$).  
 The exposure time for each individual epoch is 20--40 min., adjusted for the
 atmospheric opacity to achieve a sensitivity of 12 mJy/beam per $3^{\prime\prime}$ square pixel at
 850 $\mu$m.\footnote{This paper quotes sensitivities as pixel-to-pixel variation derived using a beam size of 14.6 arcseconds and a pixel size of 3 arcseconds, which is consistent with the sensitivities in \citet{mairs17} but may differ from methodologies in other studies.}  Each field is being observed once per month when
 available, with the first observations obtained in December 2015 and an initial program that runs through January 2019.
Since JCMT operations can extend a few hours into
dawn, each field will be observed $\sim 10$ times per year.  When all
images are stacked, the
total sensitivity at 850 $\mu$m will be $\sim 2.5$ mJy/beam \citep[compared to $\sim 4$ mJy/beam for the Gould Belt Survey,][]{mairs15}.

This monthly cadence is selected based on estimates of how quickly a luminosity burst would propagate through the envelope
and be detectable, following the radiative transfer and envelope
models calculated by \citet{johnstone13}.  Because sub-mm photons from an envelope are emitted from a large volume, the light propagation
time is a few weeks.  Once irradiated, the dust heating timescale is negligible because dust has a low heat capacity.  Therefore, a
1\~month cadence is selected as the estimated optimal cadence for 
sensitivity to accretion variability on weeks-months timescales.
This cadence will also allow us to stack several images over a few
months to characterize any smooth long term changes in the flux and to
evaluate variability of fainter objects in our field.
This stacking will reduce the flux calibration uncertainties introduced
stochastically by changes of the optical depth of the atmosphere on
timescales shorter than the integration time.

\subsection{Data Reduction and Source Extraction}

A full description of the data reduction and flux calibration is
provided in the companion paper by \citet{mairs17}, following on 
the methods developed by \citet{mairs15} and filtering on 
scales of $200^{\prime\prime}$.  
Compact peaks are measured using the JCMT Science Archive algorithm JSA\_catalogue (see the PICARD package in Starlink, \citealt{gibb13}, and Appendix A).  These peaks are not fully vetted.  Appendix A includes only those peaks that are established to be associated with a nearby disk or protostar, with a sub-mm centroid located $<7^{\prime\prime}$ from the centroid of the mid-IR peak (see Appendix A for further details).

Since flux
variability will be converted into a change in protostellar accretion
luminosity, the accuracy of our flux
calibration determines our sensitivity to accretion events.  Standard
flux calibration for sub-mm imaging with SCUBA-2 (and other similar
instruments) is calculated from contemporaneous observations of flux
standards and simultaneous measurements of the atmospheric opacity.
This standard approach to flux calibration is
accurate to $\sim 7-10\%$ \citep{dempsey13}.   

To improve upon this standard approach, we leverage the
presence of many protostars within single sub-mm fields to improve the accuracy
of our fluxes by calibrating the fluxes relative to other sources in
the same image.  A set of stable, bright peaks is identified within
each field and then used to provide the relative calibration for each
image, achieving a flux accuracy of $\sim 2-3$\% \citep{mairs17}.  This quantified uncertainty also establishes the confidence level that any detected variability is attributed to the source, rather than possible contributions because of changes in atmospheric transmission.
For faint targets, images will be stacked to look for
variability on longer timescales.

Our science results to date focus on 850 $\mu$m images.  Imaging at
450 $\mu$m is sensitive to objects in our field only during
observations obtained with low precipitable water vapor (Band 1 or
Band 2 weather, about 60\% of epochs).  Since our techniques for quantifying variability require many epochs per field, analyses of $450$ $\mu$m images will occur after we have obtained enough imaging in the best weather bands and developed techniques for the analysis of $850$ $\mu$m images.

\section{DESCRIPTION OF FIELDS AND SOURCES}

The eight star-forming regions in this survey were chosen as follows. We analyzed the JCMT Legacy Release Peak Catalogue of observed 850 micron sub-mm peaks (Graves et al. submitted) to find $30^{\prime\prime}$-wide fields with the largest number of sub-mm sources, using these as a proxy for star formation activity. Typically the brightest (most massive) sub-mm peaks in star-forming regions are found to be associated with deeply embedded protostars \citep[e.g.][]{jorgensen07,jorgensen08} and are interpreted as the molecular core out of which the star is forming. A subset of these bright (massive) peaks are not known to harbor protostars and are interpreted as being at an earlier evolutionary stage, i.e. starless or prestellar (see, e.g., \citealt{Sadavoy10} and \citealt{Pattle15,pattle17}).
As the sub-mm peaks get fainter (lower envelope mass), the association with protostars diminishes, although the mass function of the starless cores subset suggests that they may still be related to the star formation process \citep[e.g.][]{Motte98}. Many of the fainter peaks may be wispy structure within the molecular cloud and not directly related to ongoing star formation.  Some fainter emission peaks may be disks associated with Class II protostars.

Figure 2 and Table 1 show the total number of sources with peak 850 $\mu$m flux brighter than 0.125 Jy/beam for all regions in our survey, sorted by brightness bin and by association with protostars and disks (see also Appendix A).  To better understand the distribution of protostellar cores versus starless cores in our regions, we collated the sub-mm peaks against catalogues of known protostars (Class 0/I objects) and disks (Class II objects), as identified in past Spitzer
photometry \citep{dunham15,megeath16} as well as extensive sub-mm
imaging \citep[e.g.][]{johnstone99,hatchell05,kirk06,johnstone06,Enoch06,Enoch07}. 
Of the 342 bright peaks, 73 are associated with known protostars (Class 0/I) and 20 are associated with known disks (Class II).
Given the poor resolution of the JCMT, some of these associations may be coincidental, particularly in the case of disks (most of which are too faint to be detected with our sensitivity).  Moreover, since the determination of protostellar class is  often ambiguous, some sources identified as disks are more deeply embedded sources, if for example the source is viewed through a hole in the envelope.  Similarly, some of the most deeply embedded protostars are missed by the surveys due to extreme extinction, even in the mid-IR, so the lack of association of a peak with a known protostar does not rule out the presence of a protostar within the peak.  Indeed, \citet{stutz13} found a few PACS Bright Red Sources (PBRs) objects at 70 $\mu$m with Hershel that were entirely unseen with Spitzer.  Given these caveats, the numbers presented here should provide a reasonable measure of the degree of star formation activity taking place in the observed peaks in our survey, but any variability that we uncover will require more careful consideration of the individual peak and any neighboring protostar.

The following subsections describe each region in more detail.
 
\begin{figure*}[!t]
\epsscale{0.99}
\plottwo{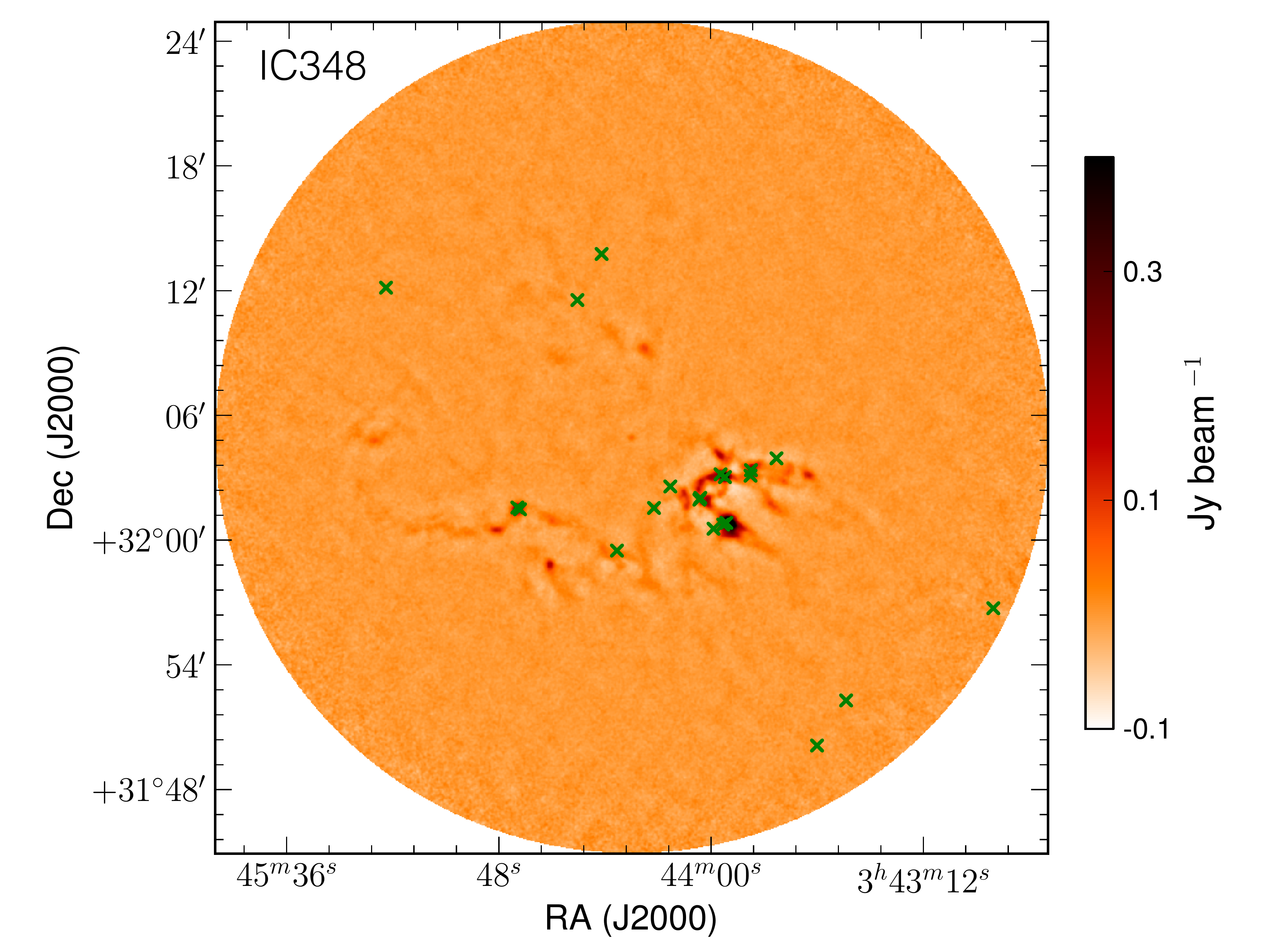}{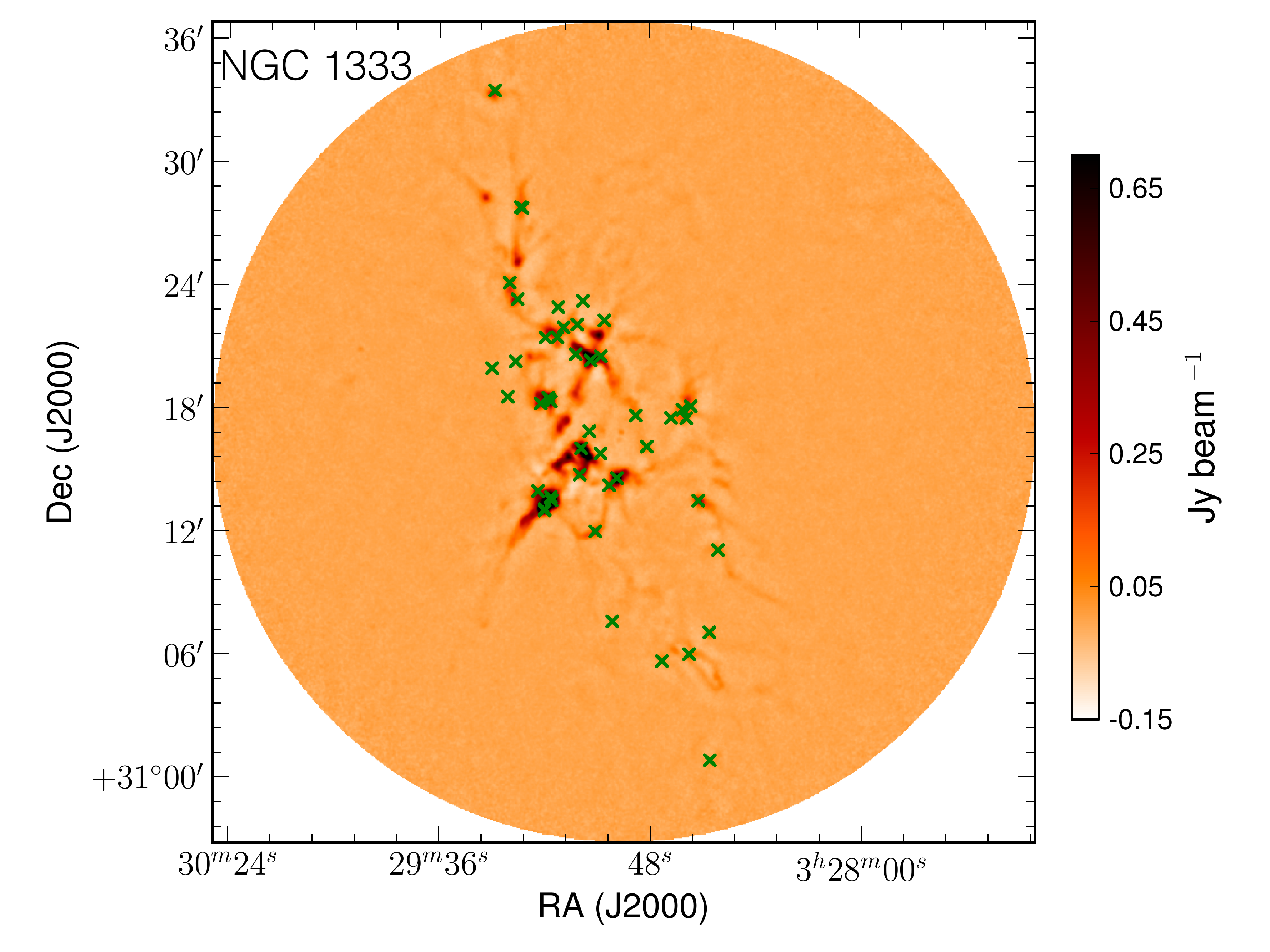}
\plottwo{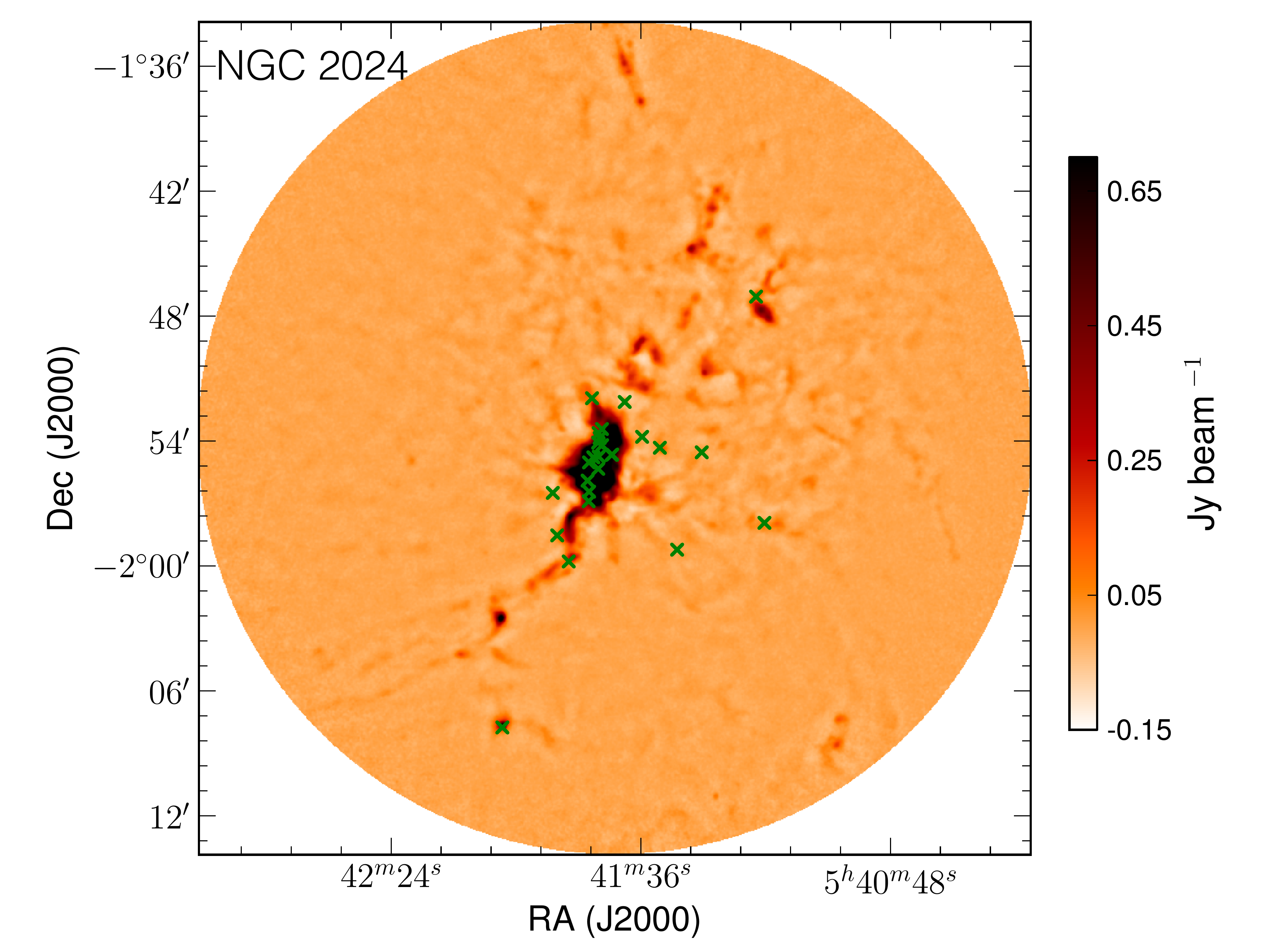}{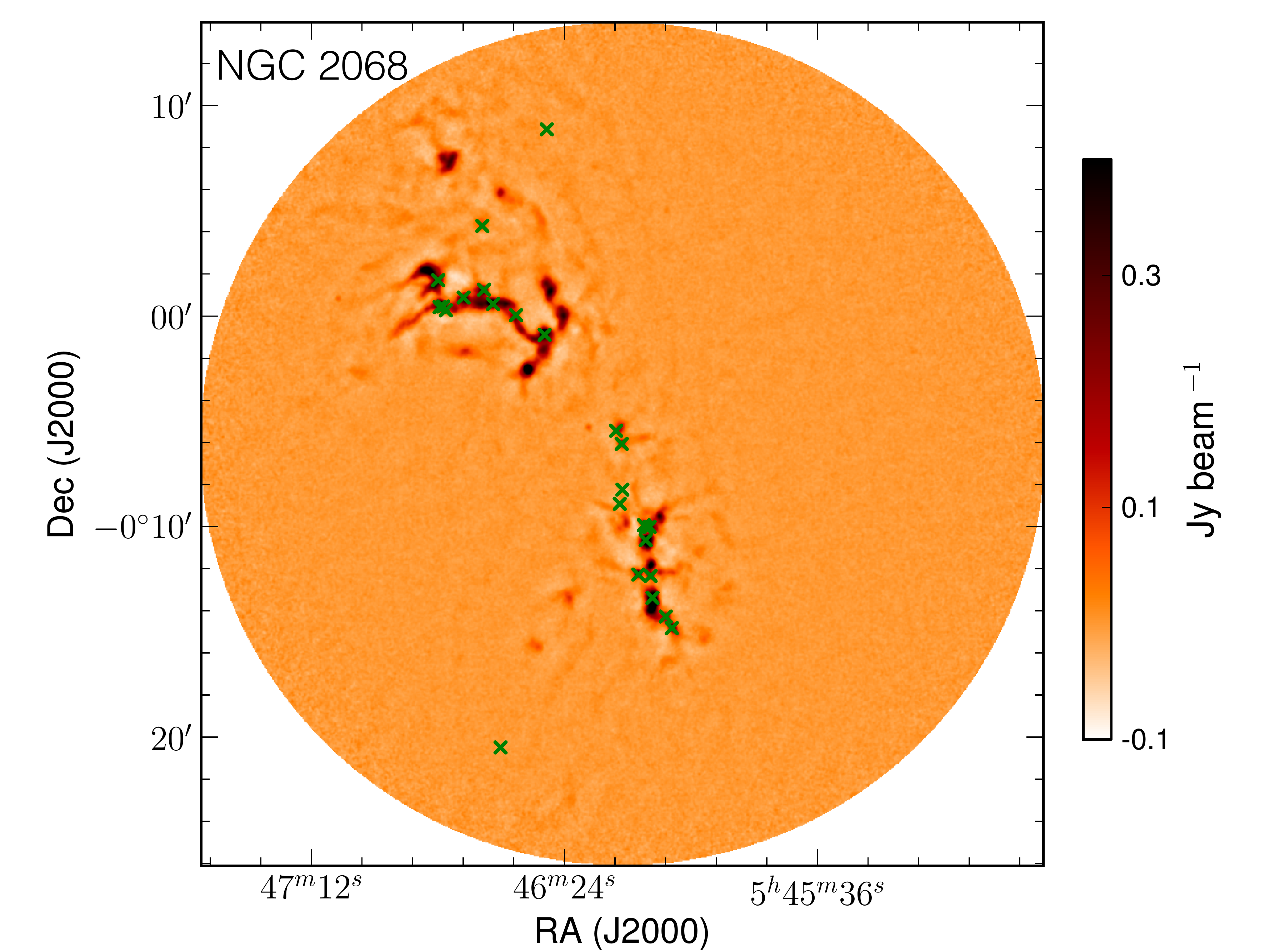}
\plottwo{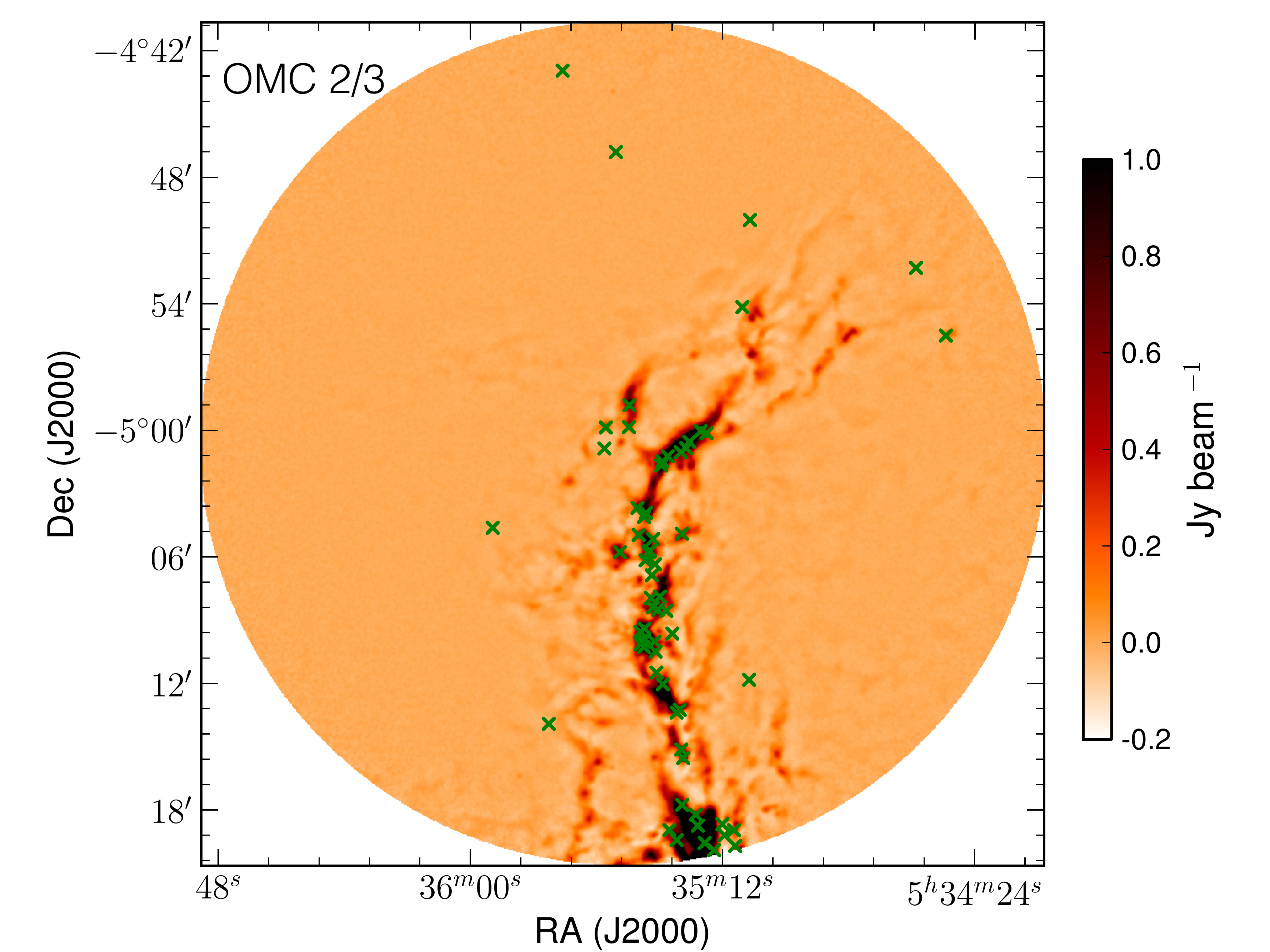}{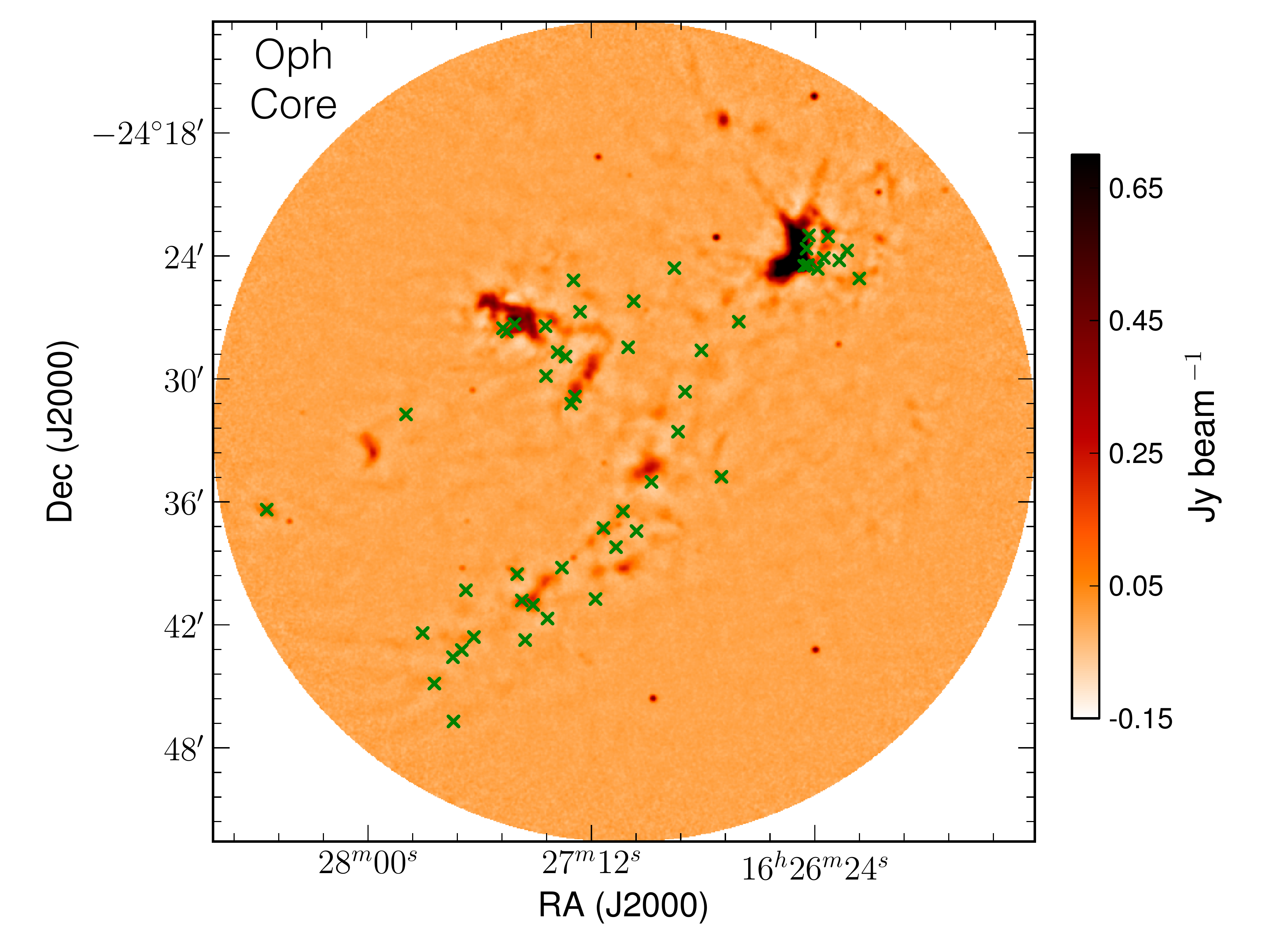}
\plottwo{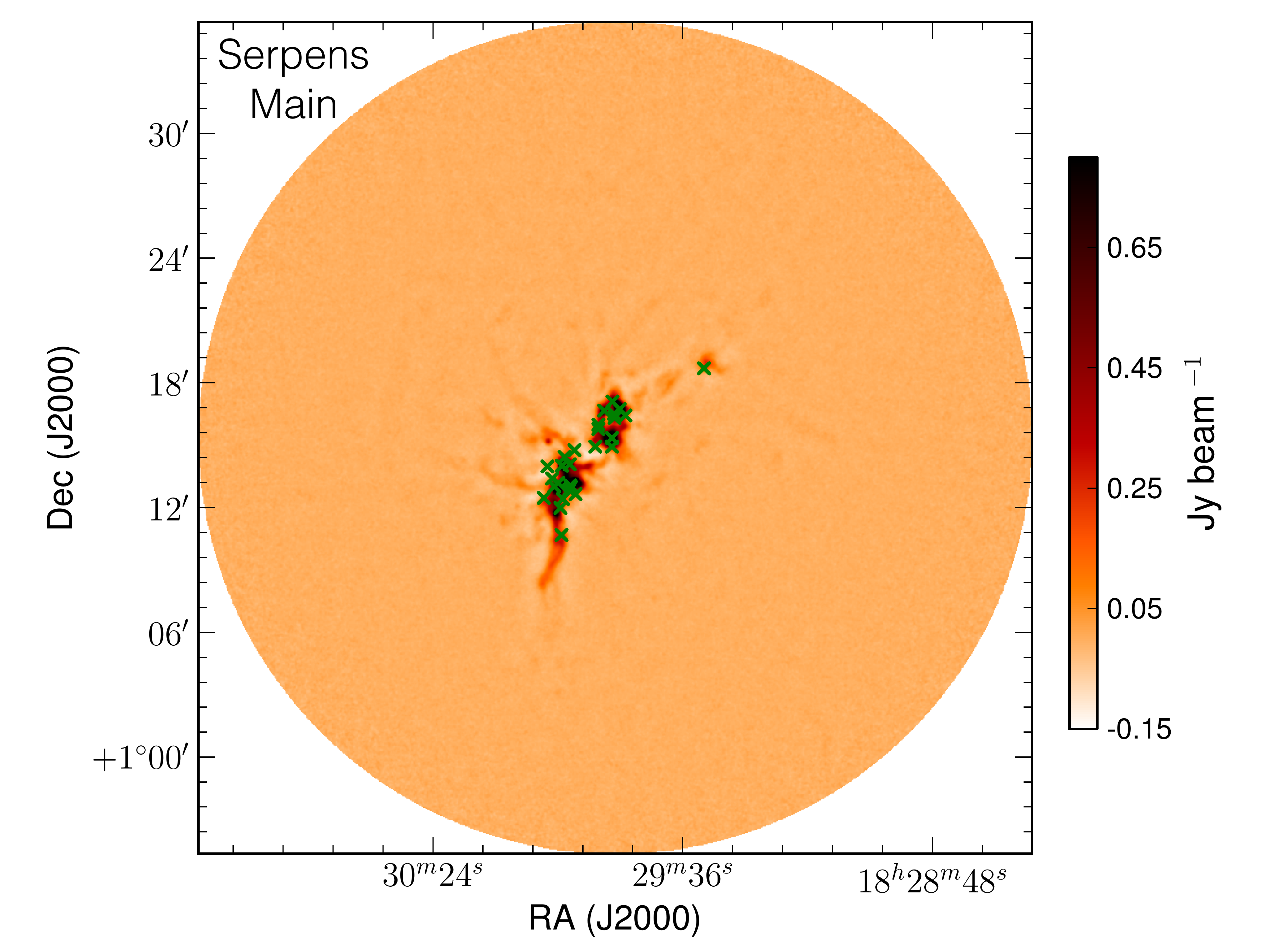}{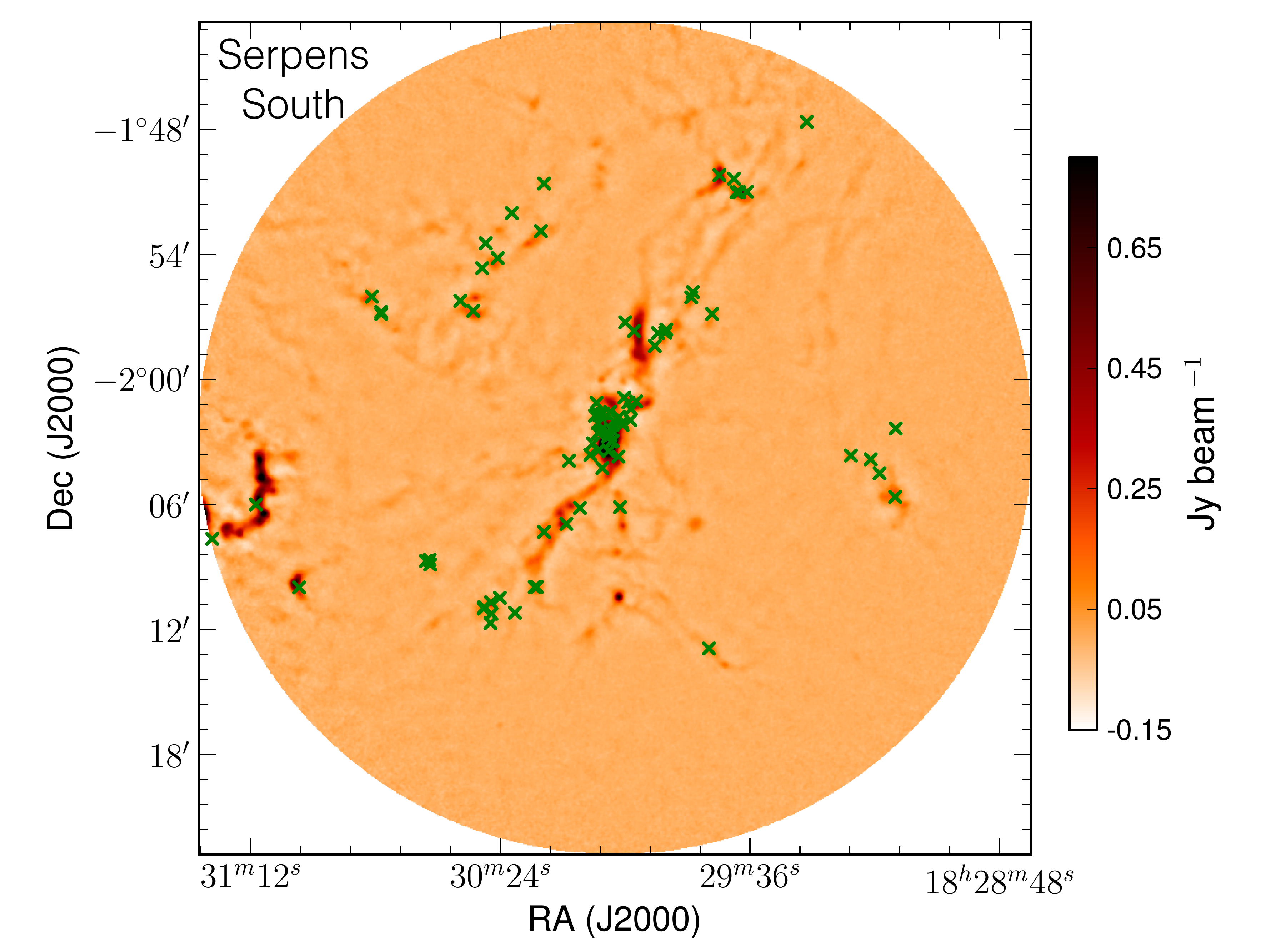}
\caption{The SCUBA-2 850 $\mu$m images of all eight regions in our survey, co-added over the first year of data \citep[see also][]{mairs17}. The marks show the location of Class 0, Class I, and flat spectrum protostars, as identified and classified by \citet{dunham15} and \citet{megeath16}.}
\label{fig:images}
\end{figure*}

\subsection{NGC 1333 -- Perseus}

NGC 1333 is in many ways a prototypical nearby star forming
cluster, with a mass of 450 M$_\odot$ and with a diameter of $\sim
1-2$ pc, located within the 
large-scale Perseus star 
forming complex \citep[e.g.][]{bally08,walawender08}. The temperatures of the regions in NGC 1333 range from 10--14 K for 
filaments and ambient cloud material, are $\sim 20$ K on the southern edge of the cloud, and reach 40~K
near the B star SVS3 and the embedded protostar IRS 6 \citep{hatchell13,chen16}.
The distance estimated from the Gaia 
TGAS catalogue \citep{gaia2016dr}\footnote{The Gaia TGAS DR1 catalogue 
contains 2 likely members within 15' of the cluster centre with distances 
of $274^{+18}_{-20}$\,pc and $267^{+19}_{-21}$\,pc \citep{gaia2016dr}.} of $271\pm20$ pc is
consistent with previous distance estimates of 220--320 pc \citep[see discussion in][]{scholz13} and would imply that NGC 1333 is located in the foreground of the Perseus star forming cloud.

Members range from 2 massive B-stars down to 
objects with estimated masses of a few Jupiter masses, with a distribution consistent with a single power law from low-mass stars to brown dwarfs  \citep{scholz12a,scholz12b,rebull14,luhman16}.
\citet{gutermuth08ngc} found that $\sim 30$\% of stars with a mid-IR excess are embedded Class I protostars, which when combined with a lack of a strong 
central condensation and localized extinction together point to an early 
evolutionary state for the cloud. The commonly-adopted age of 1--2 Myr for NGC 1333 is slightly younger than that of IC 348, although isochrone fits to optical/near-IR members 
do not show a significant age difference between the two clusters \citep{luhman16}.
NGC 1333 also  harbors several dozen
Herbig-Haro objects \citep[e.g.][]{bally96,yan98} with associated 
molecular line emission \citep{knee00,curtis10}.

\subsection{IC 348 -- Perseus}

IC 348 is a nearby star cluster associated with the Perseus cloud
complex and is located at $\sim 303$ pc from parallax
measurements.\footnote{Five members of IC348 are in the Gaia DR1 TGAS
  catalog \citep{gaia2016dr}, with an average parallax distance of
  $303\pm21$ pc, where the uncertainty is dominated by the systematic
  error of $\sim 0.3$ mas.  This distance is consistent with past
  measurements from Hipparcos \citep{Vanleeuwen07} and other methods
  \citep[see discussion in][]{herbst08}.} With an average age of 3--5
Myr \citep[e.g.][]{luhman16}, IC 348 is older than the other regions
in this survey and was selected because of the high density of
protostellar disks within the field-of-view, along with a few
protostars.  Thirteen disks were detected in previous deep SCUBA-2 images of 850 $\mu$m emission \citep{cieza15}.
Southwest of 
the main cluster is a protostellar cluster, with dense molecular clouds,
Class 0/I protostars and Herbig-Haro objects \citep[e.g.][]{walawender06}.

\subsection{OMC-2/3 -- Orion}

The Orion Molecular Cloud 2/3 region (OMC-2/3) is
located in the northern part of the Orion Molecular Cloud 
\citep[OMC,][]{bally87,mezger90} and is often referred to as the integral-shaped filament \citep{johnstone99,salji15,lane16}.  
Our pointing includes the northern half of the integral-shaped filament and contributes roughly 40\% of the total number of bright peaks in our survey.
 The OMC 2/3 region, located at \citep[$d=388$ pc from the GOBELINS VLBI parallax survey;][]{kounkel17}, is one of the 
best-studied and richest nearby star-forming regions at all observable wavelengths \citep[e.g.][]{johnstone99,tsujimoto02,megeath12} and has a disk/envelope fraction of 20\% \citep[e.g.][]{chini97,nutter07,peterson08bd,megeath12,takahashi13}.  Two sources in this region are deeply embedded PACS Bright Red Sources (PBRs) identified using far-IR photometry \citep{stutz13,tobin15}.  This region also includes (arguably) the first detected outburst of a Class 0 protostar \citep{safron15}.  Most protostars are located along 
the densest part of the molecular filaments, while Class II sources are distributed over the region.  
Sub-mm CO emission line surveys have revealed $\sim 15$ molecular outflows within this region \citep{aso00,williams03,takahashi08,takahashi12}.

\subsection{NGC 2024 -- Orion}

NGC 2024 (the Flame Nebula) is located within the Orion B (L1630) cloud complex, one of the nearest active high-mass
star-forming regions \citep[see review by][]{meyer08}, located at a distance of $\sim 388$ pc
\citep{kounkel17}.  The region is one of the densest of all clouds in Orion, with a protostellar density of $\sim 50$ pc$^{-2}$ \citep{skinner03,megeath16} spread over a virial radius of 0.4 pc \citep{lada97}.  A long molecular ridge corresponds to the regions of highest extinction and includes many dense cores \citep[e.g.][]{thronson84,visser98,choi12}.  Within the molecular ridge, two clumps of protostars are especially bright in sub-mm dust continuum emission, revealing a total mass of 633 M$_\odot$ \citep{johnstone06,kirk16}.

\subsection{NGC 2068 and HH 24 -- Orion}

NGC 2068 and HH24 (together referred to as NGC 2068 throughout this paper) are also located within the Orion B cloud complex, with HH 24 to the south of NGC 2068 \citep{kutner77,bontemps95,gibb08}.
The census of near-IR and mid-IR  protostars leads to total masses of NGC 2068 and HH 24 of
$\sim$ 240 $M_\odot$ and 120 $M_\odot$, respectively
\citep{spezzi15}. 
The protostellar population includes several PBRs \citep[e.g.][]{stutz13} and 
the eruptive
protostar V1647 Ori \citep[also known as McNeil's nebula;][]{reipurth04}.

\subsection{Ophiuchus Core}

The Ophiuchus molecular cloud spans 30 pc$^2$ on the plane of the sky and contains over 3000 solar masses of gas \citep[e.g.][]{Wilking08,dunham15}.  Star formation in the Ophiuchus complex may have been triggered by the Sco-Cen OB association, and includes numerous streamers of molecular gas pointing away from Sco-Cen (e.g., \citealt{Vrba77, Loren86, Wilking08}).  The parallax distance adopted here of 137 pc from the GOBELINS survey \citep{ortiz17ser}
is slightly larger than previous distance measurements of $\sim 120$ pc to the cluster \citep{Loinard08}.
    
The most active portion of the Ophiuchus cloud complex is L1688, which stands out from other nearby low mass star forming regions because the star forming environment is more clustered \citep{Motte98,Johnstone00b,Allen02,Johnstone04,Young06,Stanke06}. 
	L1688 shows significant substructure, with 13 identified peaks, many of which contain multiple starless and protostellar cores \citep{Loren90, Motte98, kamazaki01,Johnstone04, Young06, Stanke06, nakamura12,White15,Pan17Li}. The Oph A clump, in the northwest corner of L1688, has the highest column densities and temperatures \citep{Motte98, Johnstone00b,Friesen09, Pon09, Pan17Li}, and appears as a bright crescent of continuum emission wrapping around the position of the nearby B star S1 \citep{Elias78}.  The prototype for the class 0 protostellar stage, VLA 1623 \citep{Andre93}, is located within this field. 
    
\subsection{Serpens Main}

The Serpens Main region is an active star forming region \citep{strom76,eiroa08} with a total mass of 230--300 M$_\odot$ \citep{olmi02,graves10} located at $436\pm9$~pc \citep[from the GOBELINS survey][]{ortiz17ser}.  The SE and NW substructures are bright in sub-mm/mm dust continuum emission \citep{casali93,davis99,kaas04}.  The protostars and ongoing star formation in Serpens Main is highly concentrated at the center of our pointing.  
The sources in early evolutionary stages (Class 0/I and Flat SEDs) are clustered in small regions while the older Class II and III sources are distributed outside of these clusters \citep{harvey07}.   

The velocity field of Serpens Main shows the presence of global infall motion, outflow, rotation, and turbulence \citep{olmi02}. The velocity field in the NW subcluster is relatively uniform, on the other hand, while the SE subcluster has a more complicated velocity structure showing a large velocity dispersion ($>$0.5~km s$^{-1}$) at the central region \citep{graves10,duarte10,lee14}.  
The NW subcluster includes known IR-variables OO Ser, EC 53, and EC 37 \citep{hodapp99,kospal07,hodapp12}.  Sub-mm variability of the protostar EC 53, which was uncovered in this survey, will be presented in Yoo et al.~(2017).

\subsection{Serpens South}

Serpens South is an active star-forming region within the Aquila Rift molecular complex, located at 436 $\pm$ 9 pc \citep{ortiz17ser} and $\sim$3$\degr$ south of Serpens Main \citep{kern16}. Gas is flowing inward onto the filaments to supply the fuel for star-formation  \citep{Kirk13}.
\citet{maury11} measure a total mass of the cluster of $\sim 1660$ M$_\odot$, adjusted for the updated distance \citep[see][]{friesen16}.  
The ratio of Class 0/Class I sources of $\sim 77$\% and the Class I to Class II ratio of about 80\% are among the highest fractions for nearby star forming regions \citep{gutermuth08ser,maury11}.

\begin{figure}[!t]
\vspace{-10mm}
\epsscale{1.3}
\hspace{-14mm}
\plotone{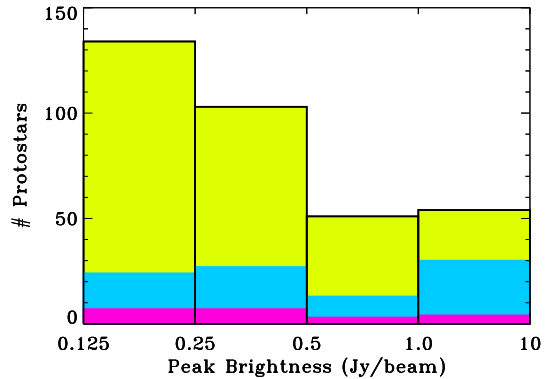}
\vspace{-55mm}
\caption{The distribution of 342 peaks with 850 $\mu$m
  peak brightness above 0.125 Jy/beam for all eight regions in our survey (yellow).  The purple and blue histograms respectively show the number of peaks associated with one or more disks and protostars.  Based on
 the analysis of \citet{mairs17}, we can achieve 2--3\% accuracy
 for the 105 peaks brighter than 0.5 Jy/beam and 10\% for the 237 sources with brightness 0.125--0.5 Jy/beam.  Of these 342 peaks, 93 are associated with distinct protostars or disks.  In some cases, multiple protostars are blended
 together to form a single peak at the resolution of JCMT.}
 \vspace{10mm}
\label{fig:peakbright}
\end{figure}

\section{TESTING MODELS OF PROTOSTELLAR ACCRETION}

During protostellar accretion, viscous processes in the disk transport angular momentum outwards, allowing gas to flow inwards towards the protostar. The source of viscosity in protostellar disks is uncertain \citep[e.g.][]{armitage15,hartmann16}; it could be due to turbulence or instabilities (gravitational, magneto-rotational) that develop where the conditions are right. When the accretion rate through some radius in the disk is lower than the accretion rate at larger radii, material builds up until the
accretion rate through the inner disk adjusts.  A steady accretion flow through the disk and onto the star is determined by the most stringent bottleneck, with short periods of strong accretion when that bottleneck breaks.  Both the amplitude of the non-steady
accretion and the timescale over which the accretion varies are likely to span a wide range of values.  Models of these accretion processes provide predictions for the
frequency and amplitude of accretion variability, with limitations related to physical scales and MHD microphysics.

\begin{figure}[!t]
\vspace{-7mm}
\hspace{-15mm}
\epsscale{1.35}
\plotone{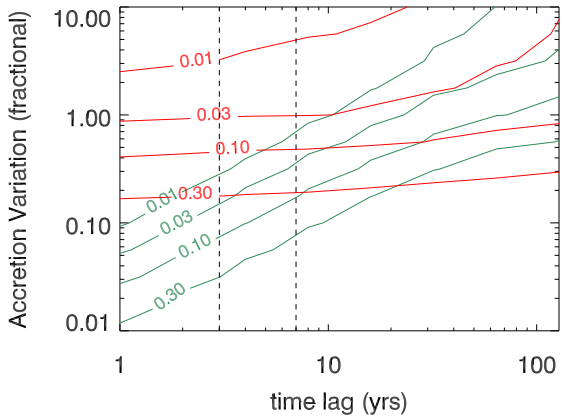}
\vspace{-8mm}
\caption{The expected fraction of time that a given theoretical model returns an amplitude variation greater than a specific amount as a function of the time lag between observations. The green contours show results for a \citet{vorobyov10} model in which accretion variability is driven by
large-scale modes within the gravitationally-unstable disk. The red contours show result for a Bae et al. (2014) model in which accretion variability is driven by the activation of the magneto-rotational instability in the otherwise magnetically inert inner disk, via heating from gravitational instability-driven spiral waves.  The contours are labeled with the fraction of stars that would show the level of variability.
In both models, larger amplitudes correlate with longer times.
The dashed lines denotes a three-year separation in time for our survey and a seven-year separation in time between earlier epochs from the JCMT Gould Belt Survey and the end of our survey.}
\label{fig:mods}
\end{figure}

Prospects for detecting accretion variability depend on the size and location of instabilities within the disk.  In the past, outbursts on young stars have been differentiated into EXors and FUors based on timescales and spectral characteristics, although it remains unclear whether these events have different physical causes or are simply different manifestations of similar phenomena \citep[e.g.][]{audard14}.
Several mechanisms have been proposed to explain short- and long-term variability, including gravitational instabilities in the outer disc region \citep{vorobyov05,vorobyov15,machida11}, thermal instabilities in the inner disc region \citep{hartmann85,lin85,bell95}, a combination of gravitational instabilities in the outer disc region and the magneto-rotational instability in the inner disc region \citep{armitage01,zhu09,zhu10}, spiral-wave instabilities \citep{bae16,hennebelle17}, gravitational interactions with companions or passing stars \citep{bonnell92,forgan10,green16,munoz16}, and magnetospheric instabilities \citep[e.g.][]{dangelo10,romanova13,armitage16}.  Some of these processes lead to the rare accretion bursts of FUors and the much more common variation of accretion seen on classical T Tauri stars \citep[e.g.][]{venuti14,costigan14,cody17}, while others may cause periodic accretion bursts, as seen in a few objects \citep{hodapp12,muzerolle13,hodapp15}.

While only some of these theoretical ideas are
capable of providing significant mass accretion variability over the lifetime of embedded protostars,
all should produce observable signatures in accretion luminosity with characteristic amplitudes
and timescales. Assuming that accretion is related to disk transport processes on orbital timescales, the
variability will depend on the radii where the physical transport processes originate and will range
from days in the inner disk to hundreds of years in the outer
disk. Within deeply embedded protostars, the range of accretion events 
taking place is almost entirely unconstrained from both
theoretical and observational perspectives.

Monitoring the brightness of deeply embedded protostars with a flux calibration uncertainty of 2--3\%  \citep{mairs17} will produce direct measurements of both the range of accretion events and their duration,
provided that the duration is longer than a few days and that the accretion is radiative and not optically-thick.  Since the total luminosity change is $\sim 10$ times larger than any change expected from the dust continuum emission at 850 $\mu$m \citep{johnstone13}, a $3$\% calibration uncertainty corresponds to a $\sim 30$\% uncertainty in the protostar luminosity.
The power
spectrum of accretion variability on young objects will then provide a
diagnostic for the size and location of disk instabilities \citep{elbakyan16},
independent of whether the majority of the mass is accreted in rare
large events.  In addition to these changes from accretion luminosity, single-epoch brightness increases may be detected from any star in our field and attributed to either short accretion bursts or non-thermal synchrotron emission (see \S 5.4).

Our monitoring of 342 peaks brighter than 0.125 mJy/beam (73 of which are associated with protostars and 20 with disks) will then provide an unbiased
survey of variability on a large enough dataset for robust statistical
analyses.   The starless cores should not be variable and therefore provide a control sample in statistical analyses.
Within our fields, 176 more protostars are fainter than 0.125 mJy/beam but increase the sample size for a search for flux increases of factors of 2--10.  While these large bursts are expected to be rare, 
models of accretion variability suggest that annual variations of
10\% may be common.  
Figure~\ref{fig:mods} presents this analysis as applied to the
outburst models of \citet{bae14} and \citet{vorobyov15}, with clear differences in the
observational signatures of accretion variability on short (less than five year) timescales that result from the different
input physics.  
In the Bae model, $>30\%$ of sources will vary by 10\%
 (our 3-$\sigma$ detection limit) over our 3.5 year program, while in
 the Vorobyov \& Basu model $\sim 7\%$ of sources would be variable
 at the 10\% level.  \citet{vorobyov15} predict that fewer bursts
 should occur during the Class 0 stage than during the Class I stage. 
 
These values suffer from large uncertainties because neither model was designed to resolve short timescales or the small distances over which the last steps of accretion occur.  However, they provide some guidance on how this program could be used as a test of models for disk accretion.
Moreover, a non-detection of
variability on this sample would indicate that the accretion flow moves smoothly
through the inner disk, placing a stringent requirement on the
instability physics in the inner disk at young ages.

Our initial investigations will search for short-term variability, as
found in EC 53 (Yoo et al., in prep), and will also place limits on the
stability of bright objects in our sample over the first year of our
program and in comparison to the Gould Belt Survey observations
obtained $\sim 3-4$ years ago (Mairs et al., in prep).
Once our survey is completed, we
will analyze all 450 and 850 $\mu$m imaging from our survey plus the
Gould Belt Survey to identify any long-term secular changes during our
$\sim 3$ years of monitoring, and with $\sim 7$ year time baselines
when including the Gould Belt Survey.

\section{RELATED SCIENCE GOALS}

After our first year of data, the stacked image from each region yields a $1\sigma$ noise level of $\sim 4$ mJy/beam at 850 $\mu$m, similar to the depth of SCUBA-2 imaging from the JCMT Gould Belt Survey \citep{mairs15} and to the deep SCUBA-2 disk surveys in the $\sigma$ Ori, 
$\lambda$ Ori, and IC~348 star-forming regions \citep{williams13,ansdell15,cieza15}.  The final stacked image after three years of monitoring should achieve a sensitivity of $\sim 2.5$ mJy/beam.
These deep images will
be useful for studying very low luminosity protostars, faint
filamentary structures, disks, and non-thermal synchrotron emission, as described below.

\subsection{Very Low Luminosity Objects (VeLLOs)}  Very Low Luminosity Objects (VeLLOs) are protostars with luminosities 
$\leq 0.1$ L$_\odot$, first discovered in {\it Spitzer} observations 
of cores thought at the time to be starless \citep[e.g.][]{young04,bourke06,dunham08,hsieh15}.
Their very low luminosities are explained by either low accretion rates or  
low protostellar masses, if they are proto-brown dwarfs
\citep[e.g.][]{lee09,lee13,hsieh16}.  The class of VeLLOs may even
include some first cores, which have short lifetimes of $10^3$ yr
before the protostar is created \citep{larson69,masunaga00}.   Using coupled disk hydrodynamics and stellar evolution models, \citet{vorobyov17}
demonstrate that the characterization of VeLLOs depends on the energy of the gas accreted onto the central object.
VeLLOs are only expected to occur if some time intervals have very low accretion rates, thereby implying other times have high accretion rates.
Our observations will reach a 
3$\sigma$ sensitivity of $\sim$ 10$^{-3}$ M$_\odot$ in envelope mass
(depending on temperature and opacity) for all regions, and should therefore be capable of unambiguously identifying 
proto-brown dwarfs in our target fields.
The factor of 2 improvement in sensitivity in our survey relative to
the Gould Belt Survey should reveal many more of these faint objects than are
detected in the Gould Belt Survey.

\subsection{Protoplanetary Disks} 

For young stars that have dispersed their envelope, a sensitivity of $\sim 2.5$\,mJy/beam
will lead to detections of disks with $\sim 1$ Earth mass
of dust in Ophiuchus and 10 Earth masses
in the more distant Orion region, assuming standard conversions between sub-mm emission and dust mass \citep[e.g.][]{andrews13}.  
Large area SCUBA-2 maps of similar depth in 
the older $\sigma$\,Ori and $\lambda$\,Ori regions have revealed that
most infrared-Class II disks have very low masses at 3--5 Myr
\citep[]{williams13,ansdell15}, and constrain planet
formation timescales more strongly than infrared surveys; similar results have been obtained from recent ALMA surveys \citep{ansdell16,barenfeld16,pascucci16} and from younger regions in the Gould Belt Survey \citep[e.g.][]{dodds15,buckle15}.  Our unbiased search for disks in some of the youngest regions of nearby star-formation
will complement the past results from older regions to establish the evolution of the disk dust mass distribution versus evolutionary stage. 

\subsection{Filamentary Structure} Much of the mass in star-forming regions is located in filamentary structures \citep[e.g.][]{andre14}.
While a full understanding requires a
combination of column density and velocity information, much can be
learned from dust continuum observations alone.  Herschel analyses
\citep[e.g.][]{arzoumanian11} have suggested that filaments 
have a characteristic width of 0.1 pc.  However, the filament width may be influenced by telescope resolution, since JCMT Gould Belt images of the Orion A
molecular cloud revealed that many filaments are significantly
narrower than 0.1 pc \citep{salji15,panopoulou17}.  
Many filaments appear to have
significant substructure along both their long and short axes and may be bundles of sub-filaments or fibers, which
have been rarely analyzed in detail but may hold important clues to
the stability and nature of the filaments \citep{,contreras13,Hacar13,hacar17}. 
Deeper SCUBA-2 observations with JCMT resolution will better reveal
faint extended substructure, thereby extending
the range of filaments to those that are lower mass and less dense, and
will allow for a 
robust measurement of filament widths.  Extended structures on scales of $>600^{\prime\prime}$ are filtered out during data reduction to account for atmospheric changes during the observations.

The filamentary structure obtained with our deep co-added integrations will also be compared with the orientation of outflows within these regions to examine the orientation of  disks in filamentary environment. 
Recent results have shown axisymmetric flattened envelopes around Class 0 sources \citep{lee12}, while outflows are often seen perpendicular to the direction of the filament \citep[e.g.][]{tobin11}. Statistics for outflows emanating from protostars that are still within their birth filamentary structures will test whether these expectations are correct, as a way to constrain the angular momentum evolution of protostars.

\subsection{Non-thermal Emission from Young Stars}

Young stars are magnetically active, producing X-rays and synchrotron
emission with a steady and a time-variable component \citep{gudel02,forbrich17}.  Flares at mm and cm wavelengths are thought to be produced by either accretion variability or by high-energy events, which produce synchrotron emission from relativistic electrons gyrating in magnetic fields \citep{bower03,massi06}.   The
time-variable component appears as flares that have been seen in
several sources to wavelength as short as 3 mm
\citep[e.g.][]{bower03,bower16,salter10,kospal11}, and may extend into the sub-mm due 
 due to synchrotron self-absorption \citep{bower03,massi06}.  
In a few cases, correlated X-ray and infrared variability may suggest that variability of high-energy emission and accretion outbursts are not necessarily distinct phenomena \citep[e.g.][]{kastner06}.

This SCUBA2 monitoring program will provide important links to longer wavelengths in the emerging context of YSO variability at infrared, X-ray, and centimeter radio wavelengths \citep[e.g][]{forbrich15,forbrich17}.  Sub-mm emission flares detected from diskless stars would be directly attributable to magnetic activity, thereby providing 
constraints on the electron energy distribution and the energetics of
the excitation mechanism for these events.  Such flares may also contaminate single-epoch sub-mm flares from protostars, though the spectral index of the 450 to 850 $\mu$m emission (if observed in good weather conditions) may allow us to discriminate between a non-thermal emission flare and a brief protostellar outburst.

\section{Summary and Future Perspectives}

Our ongoing sub-mm transient search is a novel science experiment.  We are using JCMT/SCUBA-2 to monitor once per month the 450 and 850 $\mu$m emission from eight $30^{\prime}$ fields within nearby star-forming regions.  The full survey area of $1.6$ sq.~deg. includes 105 peaks at 850 $\mu$m brighter than 0.5 Jy/beam and 237 additional peaks brighter than 0.125--0.5 Jy/beam. Of these peaks, 93 are associated with distinct protostars or disks.  In addition, 176 more protostars are fainter than 0.125 Jy/beam but increase our chances of detecting large sub-mm eruptions.  The flux calibration leverages the high density of sources in each field and is now reliable to $2-3$\% for bright sources, as expected for the 12 mJy/beam sensitivity of our images \citep{mairs17}.  This sub-mm version of differential photometry allows us to confidently quantify the stability of sub-mm sources and identify any outliers.  
Our survey is the first systematic, far-IR/sub-mm transient monitoring program dedicated to evaluating the variability on
protostars on timescales longer than a year.

Protostellar outbursts are most often found in wide-field optical
transient searches, such as the Palomar Transient Factory and ASAS-SN \citep[e.g.][]{miller11,holoien14}.
Some variability studies, such as the VVV survey and YSOVar, have targeted specific regions in the near- and mid-IR \citep[e.g.][]{contreras17,cody14}.
However, these optical/near-IR searches are not sensitive to the
most embedded (youngest) objects and include brightness changes related to variability in the line-of-sight extinction \citep[e.g.][]{aspin11,hillenbrand13}, while sub-mm luminosity should change only as the result of a variation in the protostellar luminosity.

When we began this program, we were uncertain whether any embedded
young sources would show sub-mm variability.  While outbursts detected at optical/near-IR wavelengths are
rare \citep{scholz13var,hillenbrand15}, outbursts may be much more
common at younger evolutionary phases since the disks are constantly accreting from their envelopes and may need to redistribute mass to maintain stability.  However, some models of protostellar evolution predict a lack of strong outbursts in the Class 0 phase \citep[e.g.][]{vorobyov15}.  

Within our first year,
periodic sub-mm emission has already been measured from a Class I
source and will be published in a companion paper (Yoo et al.~in prep.);
other sources show potential long-term trends.
Our future efforts will establish the frequency and size of outbursts during our $3.5$ yr survey, and also by comparing our first year of data to the previous epoch of SCUBA-2 450 and 850 $\mu$m imaging from the JCMT Gould Belt Survey to extend time baselines to $\sim 7$ yr (Mairs et al.~in prep). 
By the end of our program, the range
of variability in our sample will be able to probe the scale of disk
instabilities relevant on months- and years-long timescales.

\section{Acknowledgements}
We thank the JCMT staff for their care and attention in implementing these time-sensitive observations and for help in the reduction and distribution of the data.  We also thank the anonymous referee for their careful read and comments on the manuscript.

The authors
wish to recognize and acknowledge the very significant cultural role
and reverence that the summit of Maunakea has always had within the
indigenous Hawaiian community.  We are most fortunate to have the
opportunity to conduct observations from this mountain.  

G.~Herczeg is supported by general grant 11473005 awarded by the National
Science Foundation of China. S.~Mairs was partially supported by the Natural Sciences
and Engineering Research Council (NSERC) of Canada
graduate scholarship program. D.~Johnstone is supported
by the National Research Council of Canada and by an
NSERC Discovery Grant.  J.-E. Lee was supported by the Basic Science Research
Program through the National Research Foundation of Korea (NRF)
(grant No. NRF-2015R1A2A2A01004769)
and the Korea Astronomy and Space Science Institute under the R\&D
program (Project No. 2015-1-320-18) supervised by the Ministry of Science,
ICT and Future Planning.  Partial salary support for A.~Pon was provided by a Canadian Institute for Theoretical Astrophysics (CITA)  National Fellowship. M.~Kang was supported by Basic Science Research Program through the National Research Foundation of Korea(NRF) funded by the Ministry of  Science, ICT \& Future Planning (No. NRF-2015R1C1A1A01052160).  W.~Kwon was supported by Basic Science Research Program through the National Research Foundation of Korea (NRF) funded by the Ministry of  Science, ICT \& Future Planning (NRF-2016R1C1B2013642).  C.-W.~Lee was supported by the Basic Science Research Program though the National Research Foundation of Korea (NRF) funded by the Ministry of Education, Science, and Technology (NRF-2016R1A2B4012593).  E.~Vorobyov. acknowledges support by the Austrian Science Fund (FWF) under research grant I2549-N27.  S.-P.~Lai is thankful for the support of the Ministry of Science and Technology (MoST) of Taiwan
through Grants 102-2119-M-007-004-MY3 and 105-2119-M-007-024.

The JCMT is operated by the East
Asian Observatory on behalf of The National Astronomical Observatory
of Japan, Academia Sinica Institute of Astronomy and Astrophysics, the
Korea Astronomy and Space Science Institute, the National Astronomical
Observatories of China and the Chinese Academy of Sciences (Grant
No. XDB09000000), with additional funding support from the Science and
Technology Facilities Council of the United Kingdom and participating
universities in the United Kingdom and Canada.  The identification number for the
JCMT Transient Survey under which the SCUBA-2 data
used in this paper is M16AL001. The authors thank the
JCMT staff for their support of the data collection and
reduction efforts. The starlink software is supported by
the East Asian Observatory. This research has made use
of the NASA Astrophysics Data System and the facilities of
the Canadian Astronomy Data Centre operated by the National
Research Council of Canada with the support of the
Canadian Space Agency. This research used the services of
the Canadian Advanced Network for Astronomy Research
(CANFAR) which in turn is supported by CANARIE, Compute
Canada, University of Victoria, the National Research
Council of Canada, and the Canadian Space Agency. This
research made use of APLpy, an open-source plotting package
for Python hosted at http://aplpy.github.com, and matplotlib,
a 2D plotting library for Python (Hunter 2007).


\bibliographystyle{apj}
\bibliography{ms}


\appendix

Table~\ref{tab:allsrc} lists the 850 $\mu$m peak flux per beam for bright ($>125$ mJy/beam) sources associated with protostars. To be associated with a protostar, the peak flux location of the emission source must be less than 7$^{\prime\prime}$ from the protostar position previously listed in the {\it Spitzer}/IRS and MIPS mid-IR photometric catalogues of \citet{dunham15} and \citet{megeath16}. The emission sources are identified using the JCMT Science Archive algorithm JSA\_catalogue found in Starlink’s PICARD package \citep{gibb13}, which uses the FELLWALKER routine \citep[for more information, see][]{berry15}. The peaks are then numbered in order of brightness.  Table 2 lists the 102 protostars or disks that are associated with 93 distinct bright peaks.  In some cases, more than one protostar is located near a single 850 $\mu$m peak, and in one case a peak is associated with both a protostar and a disk (counted as a protostar in the numbers presented in the main text).  Some associations between peaks and disks and protostars may be coincidental.

\begin{table*}
\caption{Protostars and disks identified with bright sub-mm peaks$^a$}
\label{tab:allsrc}
\begin{tabular}{ccccc|ccccc}
\hline
Region & ID &Peak$^a$ & Dist. ($\arcsec$)$^b$ & Name$^c$ & Region & ID &Peak$^a$ & Dist. ($\arcsec$)$^b$ & Name$^c$\\
\hline
\multicolumn{5}{c}{Protostars} & \multicolumn{5}{c}{Protostars} \\
NGC 1333 &1 & 9.43 & 0.37 & J032910.4+311331 & Oph Core &15 & 0.35 & 2.43 & J162726.9-244050\\
NGC 1333 &2 & 3.95 & 1.92 & J032912.0+311305 & Oph Core &20 & 0.27 & 5.0 & J162709.4-243718\\
NGC 1333 &2 & 3.95 & 5.92 & J032912.0+311301 & Oph Core &24 & 0.25 & 6.19 & J162640.4-242714\\
NGC 1333 &3 & 2.88 & 6.66 & J032903.7+311603 & Oph Core &25 & 0.25 & 5.89 & J162623.5-242439\\
NGC 1333 &4 & 2.45 & 2.67 & J032855.5+311436 & Oph Core &27 & 0.23 & 1.03 & J162821.6-243623\\
NGC 1333 &5 & 1.12 & 3.09 & J032910.9+311826 & Oph Core &28 & 0.22 & 5.89 & J162727.9-243933\\
NGC 1333 &5 & 1.12 & 3.39 & J032911.2+311831 & Oph Core &29 & 0.21 & 3.22 & J162644.1-243448\\
NGC 1333 &9 & 0.59 & 3.63 & J032913.5+311358 & Oph Core &32 & 0.19 & 3.84 & J162705.2-243629\\
NGC 1333 &11 & 0.47 & 2.03 & J032857.3+311415 & Oph Core &37 & 0.15 & 2.93 & J162739.8-244315\\
NGC 1333 &12 & 0.46 & 2.27 & J032837.0+311330 & Oph Core &39 & 0.15 & 5.01 & J162617.2-242345\\
NGC 1333 &13 & 0.44 & 3.07 & J032904.0+311446 & Serpens Main & 1 & 6.76 & 3.16 & J182949.6+011521\\
NGC 1333 &17 & 0.37 & 2.1 & J032900.5+311200 & Serpens Main & 3 & 2.12 & 1.5 & J182948.1+011644\\
NGC 1333 &21 & 0.32 & 1.45 & J032917.1+312746 & Serpens Main & 4 & 1.78 & 2.0 & J182959.2+011401\\
NGC 1333 &21 & 0.32 & 4.25 & J032917.5+312748 & Serpens Main & 5 & 1.16 & 6.7 & J183000.7+011301\\
NGC 1333 &22 & 0.32 & 2.56 & J032907.7+312157 & Serpens Main & 7 & 1.06 & 2.5 & J182951.1+011640\\
NGC 1333 &24 & 0.27 & 6.63 & J032840.6+311756 & Serpens Main & 8 & 0.85 & 1.53 & J182957.7+011405\\
NGC 1333 &28 & 0.23 & 1.78 & J032834.5+310705 & Serpens Main & 9 & 0.81 & 4.23 & J182952.2+011547\\
IC 348 & 1 & 1.42 & 2.16 & J034356.5+320052 & Serpens Main & 15 & 0.28 & 2.54 & J182931.9+011842\\
IC 348 & 2 & 1.16 & 2.68 & J034356.8+320304 & Serpens South & 10 & 0.66 & 5.56 & J182938.1-015100\\
IC 348 & 3 & 0.57 & 2.64 & J034443.9+320136 & Serpens South & 17 & 0.45 & 3.0 & J183025.8-021042\\
IC 348 & 3 & 0.57 & 6.51 & J034443.3+320131 & Serpens South & 25 & 0.3 & 6.7 & J182959.4-020106\\
IC 348 & 4 & 0.34 & 1.27 & J034350.9+320324 & Serpens South & 31 & 0.2 & 4.0 & J183001.0-020608\\
IC 348 & 9 & 0.15 & 1.14 & J034412.9+320135 & Serpens South & 32 & 0.2 & 5.74 & J183017.4-020958\\
OMC 2/3 &2 & 5.96 & 6.53 & 2293 & Serpens South & 32 & 0.2 & 5.91 & J183017.0-020958\\
OMC 2/3 &3 & 5.6 & 3.08 & 2433 & Serpens South & 34 & 0.2 & 6.46 & J183015.6-020719\\
OMC 2/3 &6 & 2.66 & 5.8 & 2302 & Serpens South & 40 & 0.16 & 6.0 & J182947.0-015548\\
OMC 2/3 &7 & 2.54 & 3.24 & 2437 & Serpens South & 42 & 0.16 & 3.98 & J182912.8-020350\\
OMC 2/3 &13 & 1.74 & 3.53 & 2369 & Serpens South & 43 & 0.16 & 2.62 & J182943.9-021255\\
OMC 2/3 &13 & 1.74 & 6.85 & 2366 & Serpens South & 45 & 0.15 & 1.66 & J182943.3-015651\\
OMC 2/3 &17 & 1.58 & 3.51 & 2407 &&&&\\
OMC 2/3 &18 & 1.46 & 1.94 & 2323 & \multicolumn{5}{c}{Disks}\\
OMC 2/3 &20 & 1.39 & 1.25 & 2254 & NGC 1333 & 32 & 0.16 & 5.88 & J032856.1+311908 \\ 
OMC 2/3 &21 & 1.34 & 4.93 & 2469 & OMC 2/3 & 5 & 3.62 & 2.94 & 2072\\
OMC 2/3 &22 & 1.25 & 3.91 & 2187 & OMC 2/3 & 15 & 1.63 & 6.6 & 2334\\
OMC 2/3 &76 & 0.26 & 5.2 & 2456 & OMC 2/3 & 19 & 1.45 & 3.98 & 2029\\
OMC 2/3 &86 & 0.22 & 2.6 & 2510 & OMC 2/3 & 25 & 1.06 & 4.84 & 2345\\
NGC 2068 & 1 & 2.54 & 3.79 & 3166 & OMC 2/3 & 29 & 0.78 & 6.67 & 2371\\
NGC 2068 & 5 & 1.02 & 5.12 & 3201 & OMC 2/3 & 47 & 0.45 & 1.07 & 2179\\
NGC 2068 & 5 & 1.02 & 5.78 & 3202 & OMC 2/3 & 51 & 0.41 & 1.99 & 2347\\
NGC 2068 & 6 & 0.62 & 5.72 & 3168 & OMC 2/3 & 61 & 0.32 & 3.8 & 2184\\
NGC 2068 & 6 & 0.62 & 5.76 & 3167 & OMC 2/3 & 72 & 0.28 & 6.81 & 2145\\
NGC 2068 & 7 & 0.59 & 4.92 & 3203 & OMC 2/3 & 75 & 0.26 & 5.35 & 2333\\
NGC 2068 & 8 & 0.57 & 2.94 & 3211 & OMC 2/3 & 83 & 0.23 & 4.72 & 2363\\
NGC 2068 & 11 & 0.52 & 2.25 & 3159 & OMC 2/3 & 101 & 0.19 & 6.31 & 2228\\
NGC 2068 & 13 & 0.44 & 6.25 & 3215 & NGC 2024 & 30 & 0.17 & 3.33 & 2927 \\
NGC 2068 & 22 & 0.29 & 0.83 & 3160 & Oph Core & 4 & 0.66 & 6.25 & J162624.0-241613\\
NGC 2068 & 24 & 0.29 & 1.16 & 3180 & Oph Core & 6 & 0.59 & 4.36 & J162645.0-242307\\
NGC 2024 &4 & 5.94 & 3.37 & 2955 & Oph Core & 8 & 0.47 & 6.24 & J162323.6-244314\\
NGC 2024 &6 & 0.82 & 6.71 & 2867 & Oph Core & 18 & 0.28 & 4.89 & J162610.3-242054\\
NGC 2024 &26 & 0.19 & 4.02 & 3085 & Oph Core & 36 & 0.17 & 2.96 & J162816.5-243657\\
Oph Core & 2 & 3.98 & 5.3 & J162626.4-242430 & Oph Core & 40 & 0.14 & 2.99 & J162715.8-243843\\ 
Oph Core &13 & 0.39 & 3.71 & J162730.1-242743 & Serpens South & 31 & 0.2 & 5.41 & J183001.3-020609 \\
\hline
\multicolumn{10}{l}{$^a$Jy beam$^{-1}$}\\
\multicolumn{10}{l}{$^b$Distance between centroid of sub-mm peak and mid-IR position}\\
\multicolumn{10}{l}{$^c$\citet{dunham15} and \citet{megeath16}.}\\
\end{tabular}
\end{table*}

\end{document}